\documentclass{article}

\usepackage{PRIMEarxiv}

\usepackage[utf8]{inputenc} 
\usepackage[T1]{fontenc}    
\usepackage{hyperref}       
\usepackage{url}            
\usepackage{booktabs}       
\usepackage{amsfonts}       
\usepackage{nicefrac}       
\usepackage{microtype}      
\usepackage{lipsum}
\usepackage{fancyhdr}       
\usepackage{graphicx}       
\graphicspath{{media/}}     
\usepackage[table]{xcolor}
\usepackage{algorithm} 
\usepackage{xcolor}
\usepackage{algpseudocode}

\title{Deep Reinforcement Learning for asset allocation: Reward Clipping
}

\author{
  Jiwon Kim \\
  SK Inc.(SK C\&C) \\
  \texttt{kjiwon831@gmail.com} \\
   \And
  MOON-JU KANG \\
  \texttt{sktel1020@gmail.com} \\
  \AND
  KangHun Lee \\
  SK Inc.(SK C\&C) \\
  \texttt{potter0923@gmail.com} \\
  \And
  HyungJun Moon \\
  SK Inc.(SK C\&C) \\
  \texttt{alladdgg@gmail.com} \\
  \And
  BO-KWAN JEON \\
  SK Inc.(SK C\&C) \\
  \texttt{bk\_jeon@sk.com} \\
}

\begin{document}
\maketitle

\begin{abstract}
Recently, there are many trials to apply reinforcement learning in asset allocation for earning more stable profits. In this paper, we compare performance between several reinforcement learning algorithms - actor-only, actor-critic and PPO models. Furthermore, we analyze each models' character and then introduce the advanced algorithm, so called Reward clipping model. It seems that the Reward Clipping model is better than other existing models in finance domain, especially portfolio optimization - it has strength both in bull and bear markets. Finally, we compare the performance for these models with traditional investment strategies during decreasing and increasing markets.
\end{abstract}


\pagestyle{fancy}
\section{Introduction}
In recent years, AI algorithms-deep or machine learnings are used in financial market for various fields like stock prediction, auto trading, deep hedging, \cite{Fischer18}. At the same time, passing through the bull market right after COVID-19, we are currently experiencing difficulties in dealing with market by inflation and the rising interest rate. In this situation, to get more return with less risk, asset allocation(portfolio optimization) using Robo-Advisor with reinforcement learning is in the spotlight.\\
Zhipeng Liang et al. implement three reinforcement learning algorithm - DDPG, PPO and Adversarial PG in portfolio management in \cite{LCZ18}. They showed PG algorithm outperforms URCP in China stock market. Also, in \cite{SP20}, Farzan Soleymani and Eric Paquet present a deep reinforcement learning combined with a restricted stacked autoencoder and a convolutional neural network in portfolio management. Here they apply SARSA algorithm which is enforced with a CNN. Jung hoon Kim proposed reinforcement learning to make a short position especially in downward trends of stock markets (\cite{Kim20}).\\
This paper is composed of three parts. Firstly, we conduct existing three reinforcement algorithms: actor-only, actor-critic and PPO. In several research, it is shown that PPO has potential in portfolio management, \cite{AH20}, \cite{AX22}. Especially, Amine Mohamed Aboussalah et al. provide the stability of several RL models including PPO with a cross-sectional analysis (\cite{Wooldridge09}).\\
Here, our all three models are based on policy gradient method. Note that from \cite{Sutton18} chapter13, in a policy gradient  method, the reward function is defined by

\begin{equation}
 J(\theta) = \Sigma_{s\in S}\mu^{\pi}(s)\Sigma_{a\in A}\pi_{\theta}(a|s)q_{\pi}(s,a)
\end{equation}
where $\mu^{\pi}(s) = lim_{t \rightarrow \infty}P(s_t=s|s_0,\pi_{\theta})$ the stationary distribution for the policy $\pi_{\theta}$.\\
Furthermore (see \cite{Sutton18})

\begin{equation}
\nabla_{\theta}J(\theta)\propto \mu^{\pi}(s)\Sigma_{a\in A}q_{\pi}(s,a)\nabla_{\theta}\pi_{\theta}(a|s)
\end{equation}

Our actor-critic model contains this policy gradient method, and actor-only model has the only actor part of the actor-critic model.\\
Also, from \cite{SWDRK17}, for PPO algorithm we use

\begin{equation}
\label{eqn: original clip}
L^{CLIP}(\theta) = \hat{\mathbb{E}}_t[min(r_t(\theta)\hat{A}_t,clip(r_t(\theta),1-\epsilon,1+\epsilon)\hat{A}_t)]
\end{equation}
where $\hat{\mathbb{E}}_t$ indicates the empirical average over a finite batch of samples, in an algorithm that alternates between sampling and optimization and $\hat{A}_t$ is an estimator of the advantage function at timestep $t$. And

\begin{equation}
\label{eqn: error term}
L^{CLIP'}(\theta)=\hat{\mathbb{E}}_t[L^{CLIP}(\theta)-c_1(V_{\theta}(s)-V_{target})^2+c_2H(s,\pi_{\theta})]
\end{equation}

where $c_1$ and $c_2$ are hyperparameter constants. Here, the equation(\ref{eqn: error term}) means that when applying PPO for policy (actor) as well as value (critic) functions, besides the clipped reward, the objective function is strengthened with an error term on the value estimation and an entropy term to incentivize sufficient exploration \cite{AX22}.\\
Secondly, after checking the performance of three existing RL models and analyzing the characteristics of them, we introduce the modified new model which is called Reward Clipping model. When we test three RL models, actor-only and actor-critic models show high-risk and high-return. They gain high profitability in a bull market, but also have a big loss rate during a bear market. On the other hand, PPO model moves opposite way. It shows good defensive movement when a stock market is decreasing, but it cannot get enough return when a stock market is growing. So we combine these models to get advantages only - the result model gets high return during bull market but also good defense in a bear market. For this, we use modified PPO algorithm:

\begin{equation}
\label{eqn: new clip}
L^{CLIP}(\theta)^{NEW}=\hat{\mathbb{E}}_t[min(\hat{A}_t,clip(\hat{A}_t, 1-\epsilon_1,1+\epsilon_2)]
\end{equation}

In original PPO algorithm (equation (\ref{eqn: original clip})), clipping is given for the probability ratio $r_t(\theta)=\frac{\pi_{\theta}(a_t|s_t)}{\pi_{\theta_{old}}(a_t|s_t)}$, i.e. in our case for the proportion of each asset(product). But in financial market stability is needed for advantages- return, MDD and so on, not for portions of portfolio. Furthermore, bigger return and sharpe ratio are better, we set different values $\epsilon_1$, $\epsilon_2$ saying lower and upper bounds. So we modify the clipping logic in PPO to equation (\ref{eqn: new clip}). This is more intuitive since by controlling the advantage function directly, we can get immediate effects in our rewards. And it looks that this is more fittable model in finance area. \\
Finally, we compare performance of RL models with traditional quant investment strategies - All Weather Portfolio, 6:4 (equity:bond) and equal weight rules. These results can suggest us the direction of our RL models and give necessity of use of AI models in financial portfolio optimization.\\
In the following experiments, we use two sets of products. The first set is composed of 68 products, 22 in Europe, Korea, US bond, 44 in US, Europe, Korea, Japan equity and 2 in gold. From this we can see that the RL models give us not only an optimal asset allocation but also a product selection. For the second set, we use 25 products, 16 in US and KOREA stocks, 4 in intermediate-term treasuries, 2 in long-term treasuries, 2 in commodities including REITs and gold. With the second product set, we compare the performance of RL models to ALL Weather Portfolio strategy.

\section{Existing Models}
In this section, we implement three different existing methodologies, actor-only, actor-critic and PPO in asset optimization. We show how each models work, especially in the view of returns, sharpe ratio, standard deviation and MDD.

\subsection{Construction and Experiments}

\begin{figure}[h!]
	\centering
	\includegraphics[scale=0.8]{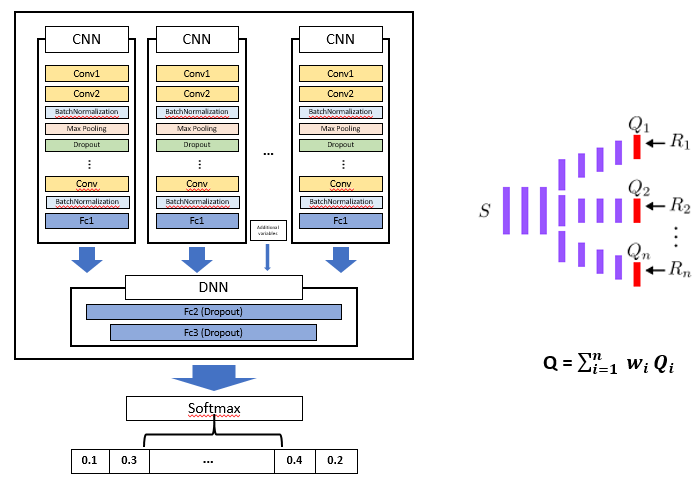}
	\caption{\label{fig:Architect} Architecture}
\end{figure}

In our experiments, one state includes previous closing price, volume or some other financial indices in a fixed window. And an action is the desired allocating weights.\\
The actor-only model is the actor part of the actor-critic model. And PPO model is the model constructed from the actor-critic model by replacing actor part to PPO algorithm. Hence all three models have the same architecture for the actor part. The following \autoref{fig:Architect} is the common architecture of three models and the output after doing softmax is the proportion for each asset.

Note that $Q$-value function is estimated using a function approximator with weight vector $\theta:Q(s,a;\theta)$ for action values. And DQN iteratively improves an estimate of $Q^*$ by minimizing the sequence of loss functions:

\begin{equation}
    \mathcal{L}_i(\theta_i)=\mathbb{E}_{s,a,r,s'}[(y_i^{DQN}-Q(s,a;\theta_i))^2],
\end{equation}

with
\begin{equation}
    y_i^{DQN}=r+\gamma {max}_{a'} Q(s',a';\theta_{i-1})
\end{equation}

Harm van Seijen et al. proposed in \cite{HRA17} to decompose the reward function $R_{env}$ into $n$ reward functions (see \autoref{fig:Architect} in \cite{HRA17}):
\begin{equation}
    R_{env}(s,a,s')=\sum^n_{k=1}R_k(s,a,s'),
\end{equation}
 for all $s,a,s'$, and to train a separate reinforcement-learning agent on each of these reward functions. Hence the associated sequence of loss function is:

\begin{equation}
    \mathcal{L}_i(\theta_i)'=\mathbb{E}_{s,a,r,s'}[\sum^n_{k=1}(y_{k,i}-Q_k(s,a;\theta_i))^2],
\end{equation}
with
\begin{equation}
    y_{k,i}=R_k(s,a,s')+\gamma \sum_{a'\in \mathcal{A}} \frac{1}{|\mathcal{A}|}Q_k(s',a';\theta_{i-1}).
\end{equation}
(See \cite{HRA17}) Here, we use return, sharpe ratio and antibias as our rewards. 

\subsection{Experimental Results}
Following is the result for three RL models: actor-only, actor-critic(AC) and PPO. We train the models from 2010-01-01 to 2019-06-10, and test them from 2019-07-18 to 2021-06-16. We want to see the movement of models for sharp drawing down and increasing stock market during the COVID-19. For this experiment, the first data set is used (a product selection of 68 products is also reflected).  

\begin{table}

	\centering
    \rowcolors{2}{gray!25}{white}
	\begin{tabular}{cccccc} 
        
        \rowcolor{gray!50}
        \vspace{0.05cm}
       {\textbf{Model}} & {\textbf{Annual Return}} & {\textbf{Sharpe Ratio}} & {\textbf{Standard Deviation}} & {\textbf{MDD}} & {\textbf{Sortino}} \\ 
        
        Actor-only  & 13.61 & 0.8068 & 0.1670 & -24.65 & 1.1432\\ [0.3EX]
		Actor-critic & 18.64 & 1.0635 & 0.1616 & -27.12 & 1.6766\\ [0.3EX]
		PPO & 10.25 & 1.0160 & 0.0966 & -18.36 & 1.4575\\ [0.5EX]
        \hline 
		
	\end{tabular}
    \vspace{0.2cm}
	\caption{\label{actor vs ac vs ppo_old} Performance of Actor-only vs AC vs PPO} 
\end{table}

\begin{figure}
	\centering
	\includegraphics[scale=0.9]{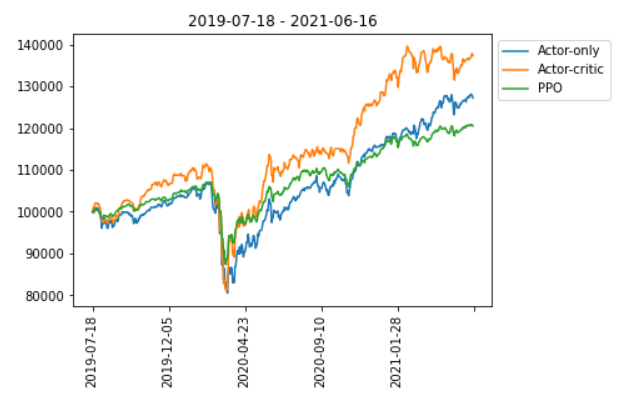}
	\caption{\label{fig:Actor-only vs AC vs PPO}Actor-only vs AC vs PPO }
\end{figure}

From the \autoref{actor vs ac vs ppo_old} and \autoref{fig:Actor-only vs AC vs PPO}, we can see that Actor-only and AC models have big draw down(MDD) but AC has good return. On the other hand, PPO model has less MDD than other two models, but small return too. Also, \autoref{fig:Actor-only vs AC vs PPO} shows some patterns for these three models as well. As we can see in the graph, general policy gradient actor(for AC and actor-only model) makes big drop. But if we compare AC and actor-only model, critic part makes the model get more returns although it cannot defense the crash. On the other hand, PPO model has smooth movement in its returns. From this, we infer that AC model works for bull-market and PPO model is good for bear market.

\section{Reward Clipping Model}
As we can see in the previous section, Actor-critic and PPO algorithm have its own characteristic. The previous one has a strength for increasing market but failing to defense in the depressed stock market. On the other hand, the PPO algorithm operates the other way around.
So here, we introduce the new algorithm so called Reward Clipping model which is strong both in increasing and decreasing stock markets.

\subsection{Idea}
\begin{figure}[h!]
    \centering
\includegraphics[scale=0.9]{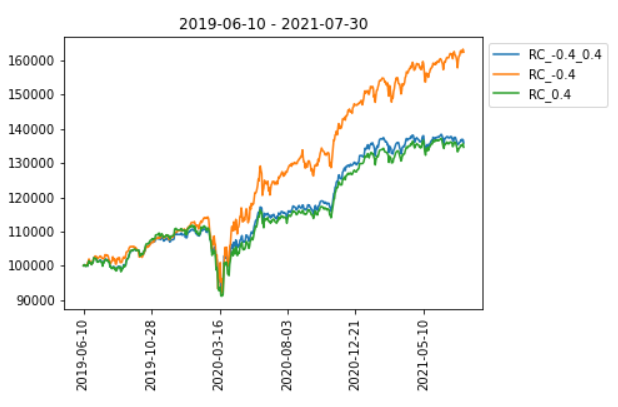}
\caption{Reward Clipping Upper and Lower bound effects}
\label{fig : Reward Clipping Upper and Lower bound effects}
\end{figure}

Note that from the PPO equation(\cite{SWDRK17}), it ensures that the update is not too large, that is the old policy is not too different from the new policy. We guess this logic makes PPO move smoothly by giving clipping on the main object, in our case proportion for each asset in a portfolio. But in a financial market, especially in an asset allocation, big changes in proportions of assets between old and new portfolio is not a problem if we get enough benefit to the point where we can ignore a turnover. Since our main purpose is return or sharp ratio even though our output is the portfolio, we apply clipping logic to our rewards, not to the main object-asset portfolio.\\
With simple experiment(not with full products) , we can see the effect of upper and lower bound in  reward clipping (see \autoref{fig : Reward Clipping Upper and Lower bound effects}). Here, RC\_-0.4\_0.4 model is the reward clipping model with both upper and lower bounds which are -0.4 and 0.4. The model RC\_0.4 is the reward clipping model with upper bound only which is 0.4. It says RC\_0.4 model has no restriction moving downward on its rewards. Similarly, RC\_-0.4 means the reward clipping has lower bound -0.4 only (no upper bound so it can move upward freely). 

As you can see the result in \autoref{fig : Reward Clipping Upper and Lower bound effects}, if models have the upper clipping bound on their rewards, it seems that they have a limitation to go up, so they cannot get enough return. On the other hands, if a model has no upper clipping bound (lower bound only, RC\_-0.4), it moves go up (gets more profit) more than other models, but less down (than other models move). Hence we can conclude that if we don't set a upper reward clipping, we can get more rewards and prevent loss of reward by giving a lower reward clipping. \\
Especially, we can find if the model has an upper bound on its reward clipping, it cannot be converged. It is because that since the model was constructed to purchase more rewards (higher reward is better), if we set up the upper bound, it seems to make confliction with the model's pursuit. The followings are the figures of the convergence for the three models. The leftmost is the convergence of RC\_-0.4\_0.4, the middle is for RC\_-0.4 and the rightmost is for RC\_0.4.

\begin{figure}[htp]

\centering
\includegraphics[width=.33\textwidth]{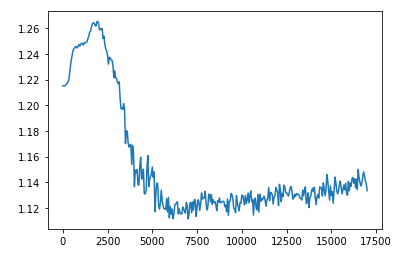}\hfill
\includegraphics[width=.33\textwidth]{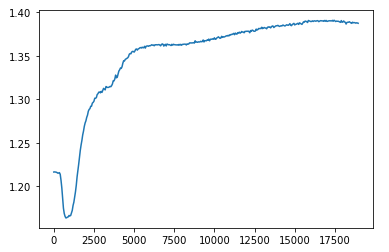}\hfill
\includegraphics[width=.33\textwidth]{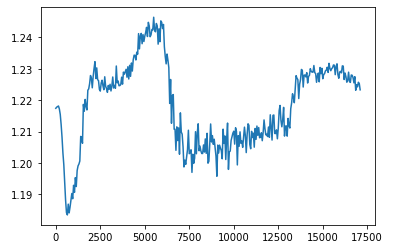}
\caption{Convergence for RC\_-0.4\_0.4, RC\_-0.4, RC\_0.4}
\label{fig:Convergence for   RC_-0.4_0.4,   RC_-0.4,   RC_0.4}

\end{figure}

\subsection{Construction}
The basic construction for the Reward Clipping model is same with \autoref{fig:Architect}. The only different part is that we apply clipping logic in PPO onto reward parts in Actor-Only model. For example, for the return reward, our formula is
$$ max (avg(\Sigma^n_{i=1} (W_i \cdot \textit{daily return}_i)))$$
where $W_i$ is the weight and $\textit{daily return}_i=(A_{i,t},...A_{i,t+T})$ and $A_{i,t}$ is the daily return at time $t$ of $i$ asset.\\ The following pseudo-codes show that which parts are modified from original PPO algorithm to Reward Clipping one.

\newpage
\begin{algorithm}[h]
	\caption{PPO-Clip} 
	\begin{algorithmic}[1]
	    
		\For {$iteration=1,2,\ldots$}
			\For {$actor=1,2,\ldots,N$}
				\State Run policy $\pi_{\theta_{old}}$ in environment for $T$ time steps
				\State Compute advantage estimates $\hat{A}_1,\ldots,\hat{A}_T$ where $\hat{A}_{t}=W_i \cdot A_{i,t}$
			\EndFor
			\State Update the policy by maximizing the PPO-Clip objective:
			$$ \theta_{k+1}={argmax}_{\theta} \frac{1}{T}\sum^T_{t=0}min(\frac{\pi_{\theta}(a_t|s_t)}{\pi_{\theta_k}(a_t|s_t)}A^{\pi_{\theta_k}}(s_t,a_t),g(\epsilon,A^{\pi_{\theta_k}}(s_t,a_t)))$$
			
			\State Optimize surrogate $L$ wrt. $\theta$, with $K$ epochs and minibatch size $M\leq NT$
			\State $\theta_{old}\leftarrow\theta$
		\EndFor
	\end{algorithmic} 
\end{algorithm}

\begin{algorithm}[h]
   
	\caption{Reward-Clip} 
	\label{reward-clip}
	\begin{algorithmic}[1]
	    
		\For {$iteration=1,2,\ldots$}
			\For {$actor=1,2,\ldots,N$}
				\State Run policy $\pi_{\theta_{old}}$ in environment for $T$ time steps
				\State Compute advantage estimates $\hat{A}_{1},\ldots,\hat{A}_{T}$ where $\hat{A}_{t}=W_i \cdot A_{i,t}$
			\EndFor
			\State Update the policy by maximizing the Reward-Clip objective:
			\begin{equation}
			\label{eqn: reward-clip}
			\color{blue} \theta_{k+1}={argmax}_{\theta} \frac{1}{T}\sum^T_{t=0}min(\frac{A_t}{A_{t-1}},\epsilon_1,\epsilon_2)
			\end{equation}
		    	where $\epsilon_1,\epsilon_2$ are lower and upper bounds.
			
			\State Optimize surrogate $L$ wrt. $\theta$, with $K$ epochs and minibatch size $M\leq NT$
			\State $\theta_{old}\leftarrow\theta$
		\EndFor
	\end{algorithmic} 
\end{algorithm}

The \autoref{eqn: reward-clip} in \autoref{reward-clip} is the biggest changed part in our new model. \\
Note that in PPO algorithm, the clipping object is the result of the action-portfolio, but the Reward-Clip object is the reward. Since we apply reward clipping to actor-only, in the above pseudo-code, the critic part is excluded.\\
With the simple experiment introduced in the previous section, we only consider the model with a lower clipping bound in its rewards.

\subsection{Experimental Results and Model Comparisons}
In the next two subsections, we give two experimental results and comparisons. To see that the reward clipping model has strength in a bear market but has enough profit in a bull market, we conduct two experiments during two period- falling and increasing markets and compare its performance with other models.

\subsubsection{Reward Clipping in a falling market}
Here, we check the effect of the Reward Clipping model in a falling market. We train the model from 2010-01-01 to 2021-06-10 and test it from 2021-07-26 to 2022-07-22. We pick this test period to see how the reward clipping model with lower bound work in current market situation. Since we apply the reward clipping logic on actor-only model, to see the effect of lower bounded reward clipping, we compare performance with actor-only model. As you can see in \autoref{fig : Reward Clipping effect in a falling market}, reward clipping with lower bound (and no upper bound) is effective for a falling market but the same return with actor-only model when a market is increasing. The detail is given in \autoref{ Table for Reward Clipping effect in a falling market}. \\
To see the market trend like the degree of decline, we put KOSPI and S\&P500 indices too. 

\newpage
\begin{table}
	
	\centering
    \rowcolors{2}{gray!25}{white}
	\begin{tabular}{cccccc} 
        
        \rowcolor{gray!50}
        \vspace{0.05cm}
	 {\textbf{Model}} & {\textbf{Annual Return}} & {\textbf{Sharpe Ratio}} & {\textbf{Standard Deviation}} & {\textbf{MDD}} & {\textbf{Sortino}} \\ [0.3EX]
		Actor-only & -6.95 & -0.3616 & 0.1633 & -20.86 & -0.5544\\ [0.3EX]
		Reward Clipping & -4.21 & -0.2809 & 0.1256 & -14.54 & -0.4422\\[0.3EX]
		KOSPI & -24.80 & -1.6585 & 0.1653 & -30.13 & -2.5470\\ [0.3EX]
		S\&P500 & -10.02 & -0.4394 & 0.1972 & -23.39 & -0.6689\\ [0.5EX]
        \hline
		
	\end{tabular}
    \vspace{0.2cm}
	\caption{\label{ Table for Reward Clipping effect in a falling market} Table for Reward Clipping effect in a falling market} 
\end{table}

\begin{figure}[h!]
    \centering
\includegraphics[scale=1.1]{image/falling_market.PNG}
\caption{Reward Clipping effect in a falling market}
\label{fig : Reward Clipping effect in a falling market}
\end{figure}

In this test(in a falling market), to compare the results with All weather portfolio (in the next section), we use the second set of products (16 products in equity, 6 in bond, 2 in commodities and 1 gold). With the above MDD and sortino (and Annual Return) in \autoref{ Table for Reward Clipping effect in a falling market}, we can see that the reward clipping model with a lower bound has a good defense in a falling market situation.

\begin{figure}[htp]

\centering
\includegraphics[width=.4\textwidth]{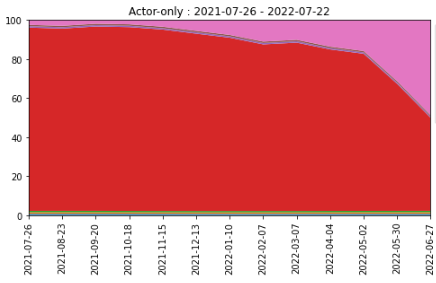}\hfill
\includegraphics[width=.5\textwidth]{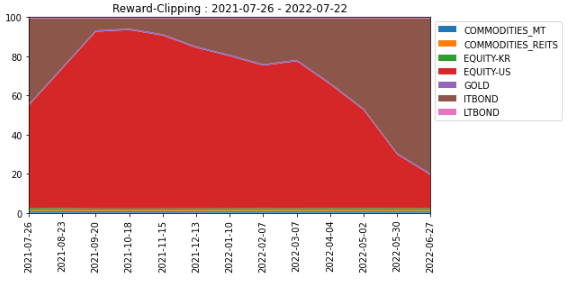}

\caption{Proportion of asset classes in bear market}
\label{fig:Proportion of asset classes in bear market}

\end{figure}

The following \autoref{fig:Proportion of asset classes in bear market} is shown the proportion of asset classes for Actor-only and RC models.

Here, we can see that RC model defense the bear market better than the Actor-only model (especially after April, 2022) by increasing the portion of Intermediate-term bond (ITBOND).

\newpage
\subsubsection{Model Comparison for four models}
\begin{figure}[h!]
    \centering
\includegraphics[scale=0.8]{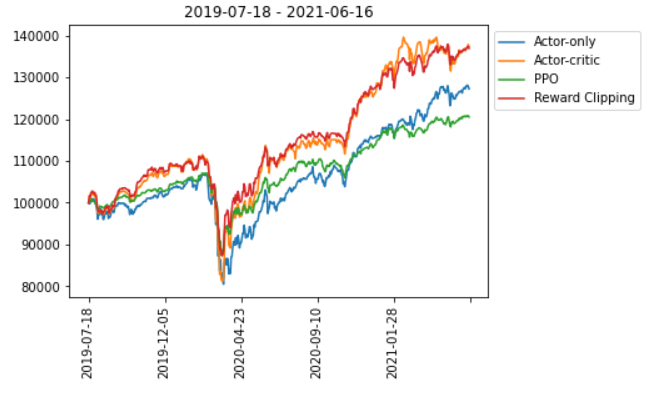}
\caption{comparison four models}
\label{fig : comparison four models}
\end{figure}

The below \autoref{comparison four models} and \autoref{fig : comparison four models} show comparison of four models- Actor-only, AC, PPO and Reward Clipping. In section 2.2 we've already seen the result for existing three models, so we just add the performance of Reward Clipping model.

\begin{table}[h]
	\centering
    \rowcolors{2}{gray!25}{white}
	\begin{tabular}{cccccc} 
        
        \rowcolor{gray!50}
        \vspace{0.05cm}
		{\textbf{Model}} & {\textbf{Annual Return}} & {\textbf{Sharpe Ratio}} & {\textbf{Standard Deviation}} & {\textbf{MDD}} & {\textbf{Sortino}} \\ [0.3EX]
		Actor-only & 13.61 & 0.8068 & 0.1670 & -24.65 & 1.1432\\ [0.3EX]
		Actor-critic & 18.64 & 1.0635 & 0.1616 & -27.12 & 1.6766\\[0.3EX]
		PPO & 10.25 & 1.0160 & 0.0966 & -18.36 & 1.4575\\[0.3EX]
		Reward Clipping & 18.45 & 1.2746 & 0.1301 & -21.45 & 2.0391\\ [0.5EX]
		\hline
	\end{tabular}
    \vspace{0.2cm}
	\caption{\label{comparison four models} Comparison four models} 
\end{table}

\begin{figure}

    \centering
    \includegraphics[width=.43\textwidth]{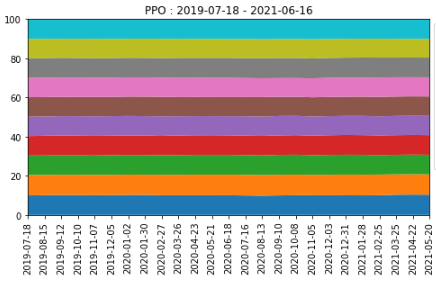}\hfill
    \includegraphics[width=.5\textwidth]{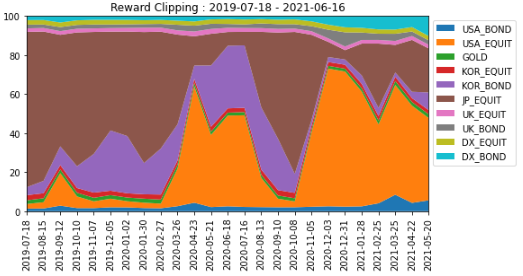}
    \caption{Proportion of asset classes in bull market}
    \label{fig:Proportion of asset classes in bull market}

\end{figure}

If you compare Actor-only and Reward clipping models in \autoref{fig : comparison four models}, we can see that Reward clipping has less draw down but more benefits in increasing situation. You can check this in \autoref{comparison four models} by comparing MDD, sortino and Annual Return - Reward clipping model has less MDD but bigger Annual Return and sortino than Actor-only model. It has the almost same bottom point to PPO but the same top point to AC at the end. It has the same increasing strength with Actor-critic but also the same defensive power with the PPO algorithm. This means by clipping onto reward in Actor-only model, we can get advantages of both Actor-critic and PPO algorithms - strength both in increasing and decreasing stock markets. 
Furthermore, as you can see in \autoref{fig:Convergence for   RC_-0.4_0.4,   RC_-0.4,   RC_0.4}, reward clipping model with lower bound doesn't much resource (actually it turns out that the reward clipping model requires less resources than PPO model) so we have benefit in the point of view of resources and time saving. \\
The following \autoref{fig:Proportion of asset classes in bull market} shows the change of proportion of asset classes. Note that we apply rebalancing every month regularly. As we can see in \autoref{fig:Proportion of asset classes in bull market} the existing three models - Actor only, Actor critic and PPO have stable movement. Especially, PPO shows almost constant movement - it is almost the same with equal weight. On the other hand, Reward Clipping model moves actively that is supposed the basis why the model has good performance in both bull and bear markets.\\
Furthermore, since PPO model needs more resources - for example, time for convergence, in the above result we can see not only the goodness of the performance but also resource effectiveness (\autoref{fig:Convergence for   RC_-0.4_0.4,   RC_-0.4,   RC_0.4}) of the reward clipping model. 

\begin{figure}[htp]

        \centering
        \includegraphics[width=.43\textwidth]{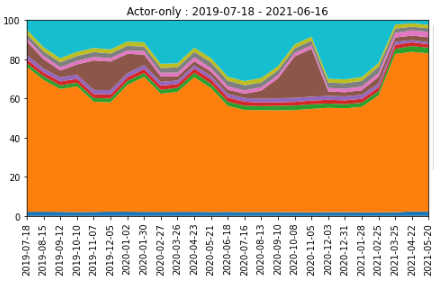}\hfill
        \includegraphics[width=.5\textwidth]{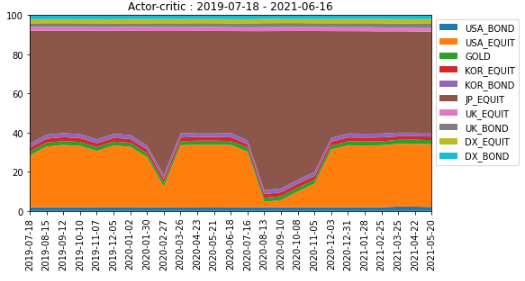}
\end{figure}  

\newpage
\section{Further work}
There are still many interesting further work using deep reinforcement learning in asset allocation. 
Firstly, we deal with ETF(Exchange Traded Fund)s only since each of them has representative index so AI models can train the indices - and so we can also contain ETF's which are launched recently although there are not enough time to train a model. But many financial corporation or customers require to expand products to several financial products - stocks, funds and so on. 
Secondly, in this paper we have applied reward clipping algorithm to actor-only model. So, the next step is to apply reward clipping algorithm to actor-critic model. Since actor-critic model has higher return than actor-only model and similar draw down, by defending the fall of actor-critic model, we expect that actor-critic model with lower bounded reward clipping has better performance. Thirdly, in our tests, we execute rebalancing every month regularly, but in a real situation, risk management system is also a necessary requisite. Actually there are several trials to apply AI to detect and react risks. In \cite{LNOM14} Yang-Yu Liu et al. show a phenomenon called "loss aversion" which says that people are much more sensitive to losses than to gains of the same magnitude. And it will affect individual decision-makings and portfolio asset prices in financial markets (\cite{CKBO17}, \cite{Yang19}). With this prior research outcomes, Qing Yang Eddy Lim et al. provide an alternative view in maximising portfolio returns using RL by considering dynamic risks appropriate to market conditions through dynamic portfolio rebalancing (\cite{LCQ22}). Finally, we can still try other RL algorithms. Although we select Actor-critic and PPO models by limitations of resources in this paper, there are many other trials to apply RL algorithms in asset allocation (\cite{AS20}). In \cite{Durall22}, Ricard Durall conduct 9 different algorithms including A2C, PPO, DDPG, SAC and TD3.  

\section{Conclusion}
\begin{figure}[ht]
\begin{minipage}[b]{.5\textwidth}
\centering
\includegraphics[width=1\textwidth]{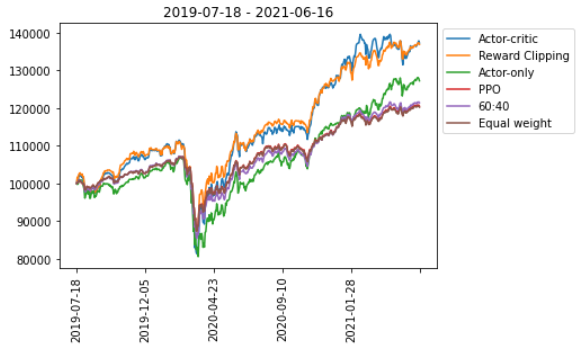}
\end{minipage}
\hfill
\begin{minipage}[b]{.5\textwidth}
\includegraphics[width=1\textwidth]{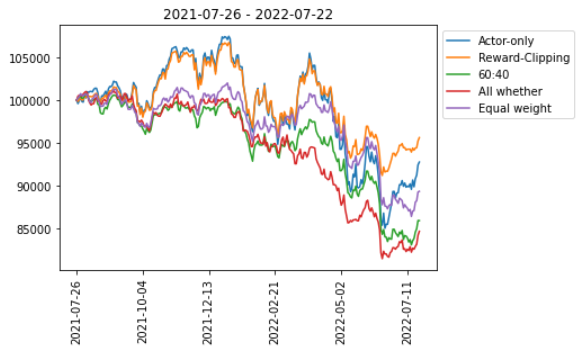}
\end{minipage}
\caption{models vs traditional approaches during COVID-19(bull market) and a bear market}
\label{fig:models vs traditional approaches}

\end{figure}

In this paper, we apply deep reinforcement learning algorithms to portfolio optimization. At first, we compare the performances of existing models-  Actor-only, Actor-critic and PPO. And then analyze the characteristics of three models. Finally, we introduce a new model which has strengths only of each models - the new model, Reward Clipping model has out-performed return in a bull market but also a good defense in a bear market. To see the model's performance, we compare them with the traditional approaches - Equal Weight, 6:4(equity:bond) and All-Weather portfolio (\cite{YL22}). Here we apply All-Weather portfolio only to the second product set (in a bear market) because the second set is consisted of proper asset classes for All-Weather method.\\
In \autoref{models vs traditional approaches during COVID-19}, \autoref{fig:models vs traditional approaches} and \autoref{ models vs traditional approaches in a bear market}, we can see that Equal weight has less MDD than other models, but small return in a bull market. When we consider Return, MDD, and sortino rate, Reward Clipping model works best for both bull and bear markets. 

\begin{table}[h]
	
	\centering
    \rowcolors{2}{gray!25}{white}
	\begin{tabular}{cccccc} 
        
        \rowcolor{gray!50}
        \vspace{0.05cm}
	
		{\textbf{Model}} & {\textbf{Annual Return}} & {\textbf{Sharpe Ratio}} & {\textbf{Standard Deviation}} & {\textbf{MDD}} & {\textbf{Sortino}} \\ [0.3EX]
		Actor-only & 13.61 & 0.8068 & 0.1670 & -24.65 & 1.1432\\ [0.3EX]
		Actor-critic & 18.64 & 1.0635 & 0.1616 & -27.12 & 1.6766\\[0.3EX]
	  PPO & 10.25 & 1.0160 & 0.0966 & -18.36 & 1.4575\\[0.3EX]
		Reward Clipping & 18.45 & 1.2746 & 0.1301 & -21.45 & 2.0391\\[0.3EX]
		Equal weight & 10.10 & 1.0012 & 0.0968 & -18.44 & 1.4386\\[0.3EX]
		6:4 & 10.70 & 0.9588 & 0.1074 & -20.36 & 1.3811 \\ [0.5EX]
    \hline
		
	\end{tabular}
    \vspace{0.2cm}
	\caption{\label{models vs traditional approaches during COVID-19} models vs traditional approaches during COVID-19} 
\end{table}

\begin{table}[h]
	 
	\centering
    \rowcolors{2}{gray!25}{white}
	\begin{tabular}{cccccc} 
        
        \rowcolor{gray!50}
        \vspace{0.05cm}
		{\textbf{Model}} & {\textbf{Annual Return}} & {\textbf{Sharpe Ratio}} & {\textbf{Standard Deviation}} & {\textbf{MDD}} & {\textbf{Sortino}} \\ [0.3EX] 
		Actor-only & -6.95 & -0.3616 & 0.1633 & -20.86 & -0.5544\\ [0.3EX] 
		Reward Clipping & -4.21 & -0.2809 & 0.1256 & -14.54 & -0.4422\\[0.3EX] 
		Equal weight & -10.23 & -1.0956 & 0.0950 & -15.27 & -1.5984 \\[0.3EX] 
		6:4 & -13.53 & -1.5420 & 0.0921 & -17.43 & -2.2477 \\[0.3EX] 
		All-Weather & -14.73 & -1.7794 & 0.0880 & -19.39 & -2.4975 \\[0.3EX] 
		KOSPI & -24.80 & -1.6585 & 0.1653 & -30.13 & -2.5470\\[0.3EX] 
		S\&P500 & -10.02 & -0.4394 & 0.1972 & -23.39 & -0.6689 \\[0.5EX] 
		
	\hline
		
	\end{tabular}
    \vspace{0.2cm}
	\caption{\label{ models vs traditional approaches in a bear market} models vs traditional approaches in a bear market}
\end{table}

In our experiments, the existing models have its own characteristics -  some have an advantage in defense for drawing down and others have a strength for profits but not in both. And also depending on the direction of each model, it seems that each model selects two or three main products/asset classes to achieve their purpose. But Reward Clipping model which the model has advantages of existing models we introduced has a strong strength for both in two opposite market situations (\autoref{models vs traditional approaches during COVID-19}, \autoref{ models vs traditional approaches in a bear market}, and \autoref{fig:models vs traditional approaches}). And it turns out that the Reward Clipping model has dynamics to select products/asset classes to pursue more profits managing a draw-down. 

\section{Acknowledgment}
We appreciate Seongjae Huh for his advice about traditional investment strategies. We also would like to say thanks to Yong Qu Lee, team leader of SK C\&C for his support.

\bibliographystyle{unsrt}  
\bibliography{references}

\begin{thebibliography}{10}

\bibitem{Fischer18}
Thomas~G. Fischer.
\newblock Reinforcement learning in financial markets - a survey.
\newblock {\em FAU Discussion Papers in Economics}, (12/2018), October 2018.

\bibitem{LCZ18}
Zipeng Liang, Hao Chen, Junhao Zhu, Kangkang Jiang, and Yanran Li.
\newblock Adversarial deep reinforcement learning in portfolio management.
\newblock {\em arXiv:1808.09940v3 [q-fin.PM]}, November 2018.

\bibitem{SP20}
Farzan Soleymani and Eric Paquet.
\newblock Financial portfolio optimization with online deep reinforcement
  learning and restricted stacked autoencoder-deepbreath.
\newblock {\em Expert Systems with Applications}, 156(113456), October 2020.

\bibitem{Kim20}
Jung hoon Kim.
\newblock Efficient portfolio management using deep reinforcement learning.
\newblock {\em Seoul National University}, December 2020.

\bibitem{AH20}
Andres Heurtas.
\newblock A reinforcement learning application for portfolio optimization in
  the stock market.
\newblock {\em UNIVERSITY OF HELSINKI}, June 2020.

\bibitem{AX22}
Amine~Mohamed Aboussalah, Ziyun Xu, and Chi-Guhn Lee.
\newblock What is the value of the cross-sectional approach to deep
  reinforcement learning?
\newblock {\em Quantitative Finance}, 22(Issue 6):1091--1111, 2022.

\bibitem{Wooldridge09}
Jeffrey~M. Wooldridge.
\newblock {\em Part 1: Regression analysis with cross sectional data.
  Introductory Econometrics A Modern Approach}.
\newblock 4th edition, 2009.

\bibitem{Sutton18}
Richard~S. Sutton and Andrew~G. Barto.
\newblock {\em Reinforcement Learning: An Introduction}.
\newblock The MIT Press, 2nd edition, 2018.

\bibitem{SWDRK17}
John Schulman, Filip Wolski, Prafulla Dhariwal, Alec Radford, and Oleg Klimov.
\newblock Proximal policy optimization algorithms.
\newblock {\em arXiv:1707.06347v2 [cs.LG]}, August 2017.

\bibitem{HRA17}
Harm van Seijen, Mehdi Fatemi, Joshua Romoff, Romain Laroche, Tavian Barnes,
  and Jeffrey Tsang.
\newblock Hybrid reward architecture for reinforcement learning.
\newblock {\em arxiv:1706.04208v2 [cs.LG]}, November 2017.

\bibitem{LNOM14}
Yang-Yu Liu, Jose~C. Nacher, Tomoshiro Ochiai, Mauro Martino, and Yaniv
  Altshuler.
\newblock Prospect theory for online financial trading.
\newblock {\em PLOS ONE}, 9(Issue 10, e109458), 2014.

\bibitem{CKBO17}
Donghyun Cheong, Young~Min Kim, Hyun~Woo Byun, Kyong~Joo Oh, and Tae~Yoon Kim.
\newblock Using genetic algorithm to support clustering-based portfolio
  optimization by investor information.
\newblock {\em Applied Soft Computing}, 61:593--602, December 2017.

\bibitem{Yang19}
Liyan Yang.
\newblock Loss aversion in financial markets.
\newblock {\em Journal of Mechanism and Institution Design}, 4(1):119--137,
  2019.

\bibitem{LCQ22}
Qing Yang~Eddy Lim, Qi~Cao, and Chai Quek.
\newblock Dynamic portfolio rebalancing through reinforcement learning.
\newblock {\em Neural Computing and Applications}, 34:7125--7139, 2022.

\bibitem{AS20}
Miquel~Noguer i~Alonso and Sonam Srivastava.
\newblock Deep reinforcement learning for asset allocation in us equities.
\newblock {\em arXiv:2010.04404v1 [q-fin.PM]}, October 2020.

\bibitem{Durall22}
Ricard Durall.
\newblock Asset allocation: From markowitz to deep reinforcement learning.
\newblock {\em arXiv:2208.07158v1 [q-fin.PM]}, July 2022.

\bibitem{YL22}
Youssef Louraoui.
\newblock The all-weather portfolio approach: The holy grail of portfolio
  management.
\newblock {\em SSRN}, (4021133), 2022.

\end{thebibliography}

\end{document}